# Skyrmion based microwave detectors and harvesting

G. Finocchio[1], M. Ricci[2], R. Tomasello[3], A. Giordano[1], M. Lanuzza[3], V. Puliafito,[1] P. Burrascano[2], B. Azzerboni[1], and M. Carpentieri[4]

[1] Department of Electronic Engineering, Industrial Chemistry and Engineering, University of Messina, c.da di Dio, I-98166, Messina, Italy

[2] Department of Engineering, Polo Scientifico Didattico di Terni, University of Perugia, Terni, TR, I-50100 Italy

[3] Department of Computer Science, Modelling, Electronics and System Science, University of Calabria, via P. Bucci 41C, I-87036, Rende (CS), Italy

[4] Department of Electrical and Information Engineering, Politecnico di Bari, via E. Orabona 4, I-70125 Bari, Italy

**Abstract.** Magnetic skyrmions are topologically protected states that are very promising for the design of the next generation of ultralow-power electronic devices. In this letter, we propose a magnetic tunnel junction based spin-transfer torque diode with a magnetic skyrmion as ground state and a perpendicular polarizer patterned as nanocontact for a local injection of the current. The key result is the possibility to achieve sensitivities (i.e. detection voltage over input microwave power) larger than 2000V/W for optimized contact diameters. Our results can be very useful for the identification of a new class of spin-torque diodes with a non-uniform ground state.

Introduction

The spin-transfer torque diode (STD) effect is a rectification effect that converts a microwave current in a dc voltage[1]. The physics at the basis of the STD effect is linked to the excitation of the ferromagnetic resonance. In particular, for a fixed input frequency, the detection voltage $V_{dc}$ is proportional to the amplitude of the microwave current, to the oscillating magneto-resistance $\Delta R_s$ and to the cosine of the phase difference between the two previous signals $\Phi_s$, $V_{dc} = \frac{1}{2} I_{ac} \Delta R_s \cos(\Phi_s)$. Since its discovery in 2005,[1] due to the low sensitivity equal to 1.4 V/W (rectified voltage over input microwave power), this effect has been used only to estimate the torques in MTJs,[2,3] although calculations from Ref. 4 suggested that optimized MTJs should reach sensitivities exceeding 10 000 V/W.

From an experimental point of view, the STD sensitivity has been improved by a simultaneous application of a microwave together with a bias current.[5,6,7] In this case, an additional component proportional to the dc current can contribute to increase the detection voltage.[8,9] Considering the advances in the design of the MTJs, in terms of tunneling magnetoresistive (TMR) effect and voltage controlled perpendicular anisotropy (VCMA)[10], and the understanding of strong non-linear effects, such as stochastic resonance,[9] non-linear resonance[6] and injection locking,[7] the sensitivity performance of biased STD has reached values as large as 75.000 V/W.[7] Those results on biased STDs open a path for the design of a new generation of high sensitivity microwave detectors. On the other hand, unbiased STDs could be also tuned to rectify the microwave power from different energy sources, including satellite signals, sound signals, television signals and Wi-Fi signals. In other words, unbiased STDs can be used for power harvesting functionality by converting a microwave signal to a dc voltage capable to recharge a cell phone battery, or to supply self-powered sensors such as other small electronic devices. Up to now, unbiased STDs have been designed with a uniform ground state and thanks to the use of both VCMA and spin-transfer torque, a sensitivity of 900 V/W has been measured.[7, 10]

Very recently, the stabilization of Néel skyrmions at room temperature has been demonstrated by several groups independently.[11, 12] This experimental milestone in the development of ultra low power devices has motivated the study presented in this letter. We report the results of micromagnetic simulations of STDs where a Neél skyrmion is the magnetic ground state of the free layer. We have considered the diode effect of the skyrmion based STD as response to a microwave current locally injected into the ferromagnet via a nano-contact.

The key quantitative result of this letter is the prediction of sensitivities reaching 2000V/W at zero field, zero bias current and low input microwave power (<0.5μW). Our finding are interesting from a technological point of view, skyrmion based STD can be the basis in the design of passive nanocontact microwave detectors or a new type of energy harvesting.[13, 14]

Device and model.

We have studied an MTJ (see sketch in Fig. 1), with a circular cross section of diameter $l$=100nm. The stack is composed of an iron-rich free layer (0.8nm) and an MgO spacer with on top a nano-contact of CoFeB acting as current polarizer of diameter $d_C$<$l$. The system is designed in order that the interfacial perpendicular anisotropy (IPA), due to the electrostatic interaction between the Fe of CoFeB and the O of MgO is large enough to induce an out of plane easy axis (z-axis) in both ferromagnets.[15] Furthermore, the free layer is coupled to a Pt layer to introduce in its energy landscape an additional degree of freedom due to the interfacial Dzyaloshinskii-Moriya Interaction (i-DMI) given by:

$$\varepsilon_{DMI} = 2D\left[m_z \nabla \cdot \mathbf{m} - (\mathbf{m} \cdot \nabla) m_z \right] \quad (1)$$

where $\mathbf{m}$ is the normalized magnetization vector of the free layer and $m_z$ is its z-component.[16] Basically, the effective field of the free layer takes into account together with the standard micromagnetic contributions of the exchange and the self-magnetostatic fields, the IPA and the i-DMI fields. All the results discussed in this letter are achieved at zero external field. The main simulation parameters are saturation magnetization $M_S$=900kA/m, Gilbert damping $\alpha_G$=0.03, and exchange constant $A$=20pJ/m. We have performed micromagnetic simulations based on the Landau-Lifshitz-Gilbert (LLG) equation to study the stability and the dynamical response of a single Néel skyrmion as a function of the IPA ($k_U$ represents the IPA constant) and of the i-DMI parameter $D$ (see Refs. 17 and 18 for a complete numerical description of the model).

The inset of Fig. 2(a) represents a snapshot of a skyrmion (the arrows refer to the in-plane component of the magnetization, while the color is linked to its out-of-plane component - blue negative and red positive) as ground state of the free layer, where also its diameter $d_{SK}$ is indicated (the radius computed as the distance from the geometrical center of the skyrmion to the region where the out-of-plane component of the magnetization $m_z$ is 0). The main panel of Fig.2(a) shows the profile of $m_z$ as computed by considering the section AA'. The negative region coincides with the skyrmion core while the non monotonic behavior near the edges is due to the boundary conditions in presence of the i-DMI $\frac{d\mathbf{m}}{dn} = \frac{1}{\xi}(\hat{z}\times\mathbf{n})\times\mathbf{m}$ ($\xi = \frac{2A}{D}$ is a characteristic length in presence of DMI and $\mathbf{n}$ is the direction normal to the surface).[19] For the nucleation of the skyrmion, we apply a localized dc spin-polarized current following the same procedure as described in Ref. 20 finding that nucleation in a time smaller than 5ns is achieved with bias current $J = 3\times10^7 \text{A/cm}^2$.

Figure 2(b) and (c) show the skyrmion diameter $d_{SK}$ as a function of $D$ for two values of $k_U$ = 0.8 and 0.9MJ/m$^3$ and as a function of $k_U$ for $D$=3.0 and 3.5 mJ/m$^2$ respectively. In our study, we have used DMI parameter larger than the one estimated in *state of the art materials* (>2.0 mJ/m$^2$ for a ferromagnetic thickness of 0.8nm).[21, 22] However, *ab initio* computations have predicted values of $D$ close to 3.0mJ/m$^2$ for Pt/Co bilayers.[23] In addition, this choice can permit to have a comparison with previous numerical studies.[24] The fixed layer diameter $d_C$, which also corresponds to the nano-contact size, has been chosen to be comparable or larger than the skyrmion diameter. This aspect is important to design the skyrmion based microwave detector in order to optimize its sensitivity as it will be discussed below. To study the microwave dynamical properties, we have considered both the Slonczewski and the field-like torque as additional terms to the LLG equation:

$$\frac{g}{|e|\gamma_0}\frac{|\mu_B|J}{M_s^2 t}g_T(\mathbf{m},\mathbf{m_p})\left[\mathbf{m}\times(\mathbf{m}\times\mathbf{m_p}) - q(V)(\mathbf{m}\times\mathbf{m_p})\right] \qquad (2)$$

where $g$ is the gyromagnetic splitting factor, $\gamma_0$ is the gyromagnetic ratio, $\mu_B$ is the Bohr magneton, $q(V)$ is a term which takes into account the voltage-dependence of the field like torque (see Ref. 25 for more numerical details), $J = J_M \sin(\omega t)$ is the microwave current density, $t$ is the thickness of the free layer, $e$ is the electron charge, and $\mathbf{m_p}$ is the normalized magnetization of the polarizer fixed along the positive z-axis. $g_T(\mathbf{m},\mathbf{m_p}) = 2\eta_T(1+\eta_T^2 \mathbf{m}\cdot\mathbf{m_p})^{-1}$ is the polarization function.[26, 27] We have used for the spin-polarization $\eta_T$ the value 0.66.[2] The resistance of the device R depends on the skyrmion diameter $d_{SK}$ as

$$R = \begin{cases} R_P\left(1 - \frac{d_{SK}^2}{d_C^2}\right) + R_{AP}\frac{d_{SK}^2}{d_C^2}; & d_{SK} < d_C \\ R_{AP}; & d_{SK} \geq d_C \end{cases} \qquad (3)$$

being $R_{AP}$ = 1500 Ω and $R_P$ =1000 Ω the resistance in the antiparallel and parallel state.[28] To characterize the dynamical response of the STD, we have also used the detection sensitivity $\varepsilon$ computed as the ratio between the detection voltage and the input microwave power ($P_{in}$) $\varepsilon = \frac{V_{dc}}{P_{in}}$. $P_{in}$ is the active power delivered to the MTJ is computed as $P_{in} = 0.5 J_M^2 S^2 R$, being $R$ the static resistance (see Eq. 3) and $S$ the contact area.

Results and Discussion

Figures 3(a) and (b) show the STD responses as a function of the microwave frequency ($d_C$=40nm) achieved for $J_M$ =3x10$^6$A/cm$^2$, in (a) $k_U$=0.8MJ/m$^3$ is maintained fixed while $D$=2.5, 3.0, 3.5 and 4.0mJ/m$^2$, in (b) $D$=3.0mJ/m$^2$ while $k_U$ changes from 0.7MJ/m$^3$ to 0.9kJ/m$^3$. The skyrmion response is mainly characterized by the excitation of a breathing mode of its core, with a preserved radial symmetry, similarly to the fundamental mode describe in Ref. 24. This mode induces an oscillation resistance of amplitude $\Delta R_s$ linked to the minimum and maximum skyrmion core diameter $d_{SK-min}$ and $d_{SK-max}$ and given by

$$\Delta R_S = \begin{cases} (R_{AP}-R_P)\dfrac{d^2_{SK-max}-d^2_{SK-min}}{d^2_C} & d_{SK-max} < d_C \\ (R_{AP}-R_P)\left(1-\dfrac{d^2_{SK-min}}{d^2_C}\right) & d_{SK-max} \geq d_C \end{cases} \quad (4)$$

$\Delta R_S$ is as larger as $d^2_{SK-min} -> 0$ and $d_{SK-max} -> d_C$.

Moreover, the ferromagnetic resonance frequency, as expected does not change with the contact diameter $d_C$ (not shown) being linked to the excitation of the fundamental mode of the skyrmion. An example of the time and space domain evolution of the skyrmion core at the resonance frequency $f$=4.6GHz ($k_U$ = 0.8MJ/m$^3$, and $D$=3.0mJ/m$^2$) is shown in Supplementary Video 1.

The ferromagnetic resonance (FMR) frequencies as a function of $D$ for different $k_U$ and as a function of $k_U$ for different $D$ are summarized in Figs. 3(c) and (d). Depending on $k_U$ and $D$, the following two regions can be identified:

$$\begin{cases} D-4\sqrt{A(k_U-0.5\mu_0 M_S^2)}/\pi < 0 & (A) \\ D-4\sqrt{A(k_U-0.5\mu_0 M_S^2)}/\pi \geq 0 & (B) \end{cases} \quad (5)$$

being $D_{crit} = 4\sqrt{A(k_U-0.5\mu_0 M_S^2)}/\pi$ the critical D (see Ref. 19 for a detailed discussion on $D_{crit}$). Figs. 3(c) and 3(d) also show different tunability of the FMR frequency as function of different parameters. This behavior has been also observed in Fig. 5(a) of Ref. 24 (see for example an ideal line at zero field in that figure). We argue this different tunability is due to the different skyrmion energy landscape near the $D_{crit}$ as extensively described in Ref. 19. In particular, for the parameters in Fig. 3(c) and 3(d) the critical values result at $k_U$ =0.8 MJ/m$^3$ $D_{crit}$ =3.07 mJ/m$^2$, at $k_U$ =0.9 MJ/m$^3$ $D_{crit}$ =3.56 mJ/m$^2$, at $D$=3.0 mJ/m$^2$ $k_{U\ -crit}$ =0.79 MJ/m$^3$ and at $D$=3.5 mJ/m$^2$ $k_{U\ crit}$ =0.89 MJ/m$^3$;

To compare the skyrmion-based STD to the state of the art MTJ-based STDs, we have performed a systematic study of the sensitivity computed at the FMR frequency as a function of the $d_C$ (30nm< $d_C$ <60nm). All the data discussed below are obtained for $k_U$ = 0.8MJ/m$^3$ and $D$=3mJ/m$^2$, however qualitative similar results have been achieved for $k_U$ = 0.9MJ/m$^3$ and $D$=2.5 and 3.5 mJ/m$^2$.

Fig. 4(a) summarizes the sensitivities computed for four different current amplitudes $J_M$ =1-4x10$^6$A/cm$^2$ maintained constant at different $d_C$. One key finding is the existence of an optimal contact size where the sensitivity exhibits a maximum value. This result can be qualitatively understood as follow. For a fixed skyrmion ground state, the change in the $d_C$ introduces a change in the microwave power (static resistance $R$ and contact cross section depend on $d_C$). Fig. 4b displays the detection voltage and the input microwave power as a function of the contact size for $J_M$ =1x10$^6$A/cm$^2$. As can be observed, the detection voltage increases while the input power decreases as $d_C$ increases. A trade off between those conditions determines the optimal $d_C$ as reported in Fig. 4a. The second result is the prediction of sensitivities of the order of 2000V/W for an optimal configuration, values which are larger than the one of *state of the art* unbiased MTJ-based STD around 900 V/W.

Fig. 4(c) summarizes the sensitivities computed as a function of the $d_C$ while maintaining a constant input microwave power. Those computations also show the existence of an optimal contact which coincides with a the maximum of the detection voltage as summarized in Fig. 4(d).

The last part of the letter investigates a possible contribution of the VCMA to the sensitivity of skyrmion based STDs.[29] We have implemented this contribution as an additive field to the effective field as $H_{VCMA} = \Delta H_{VCMA} \sin(\omega t)$ applied along the out-of-plane direction.[30] In CoFeB/MgO/CoFeB stacks, VCMA has been already used to demonstrate the electrical field assisted switching,[31] and to improve the sensitivity of unbiased STDs.[7,10] Considering the optimal scenario of Fig. 4(a) ($d_C$=40nm), we have performed an ideal numerical experiment by computing the sensitivities as a function of the VCMA, as summarized in Fig. 5, for $J_M$ =1 and 2x10$^6$A/cm$^2$ while maintaining constant the other parameters. A deep analysis of the micromagnetic configurations show that at low VCMA both $d_{SK-\min}$ and $d_{SK-\max}$ change ($d_{SK-\min}$ decreases $d_{SK-\max}$ increases), as the VCMA increases at a certain value the change in the $d_{SK-\max}$ will saturate to $d_C$ while $d_{SK-\min}$ will continue to decrease.

To estimate the VCMA contribution in our devices, we consider the optimal case of Fig. 4(a), $J_M$ =1x10$^6$A/cm$^2$ and a typical experimental VCMA value of 60mT/V[10]. The corresponding voltage change across the tunnel barrier is $\Delta V$ <15mV that gives rise to a negligible VCMA contribution $\approx 1$ mT.

In summary, we have proposed a zero field and an unbiased skyrmion based STD for passive microwave detection and energy harvesting application. The skyrmion is stabilized by the i-DMI and when it is supplied by a microwave spin current with a perpendicular polarization, a breathing mode is excited. The change in the size of the skyrmion is converted to a change of the TMR signal which gives rise to the detection voltage. We have found that sensitivities as large as 2000V/W can be achieved for optimized contact diameter. Our results open a new path for the design of STD with a non-uniform ground state.

Acknowledgments

This work was supported by the project PRIN2010ECA8P3 from Italian MIUR. The authors thank Domenico Romolo for the support in making Figure 1. MR and PB acknowledge financial support from Fondazione CARIT - Progetto Sensori Spintronici.

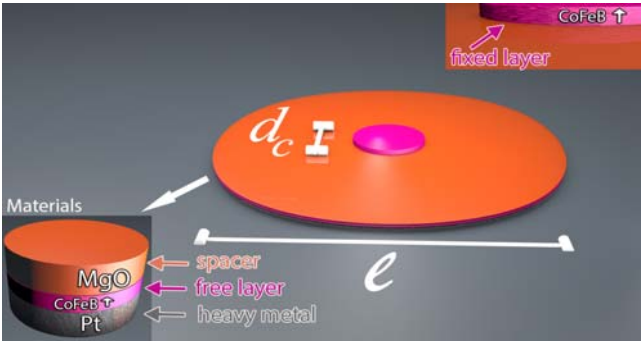

Figure. 1. Sketch of the device under investigation, the extended CoFeB acts as free layer while the top nano-contact made of CoFeB is the current polarizer ($d_C$ is the contact diameter, $l$=100nm), in the Pt layer is necessary to introduce the i-DMI.

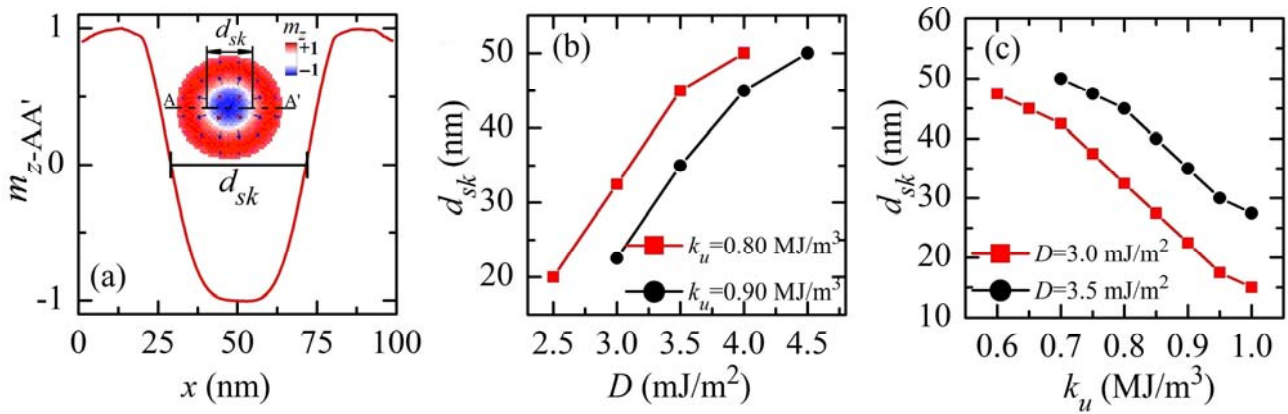

Figure 2: (a) Profile of the out-of plane component of the magnetization corresponding to the section AA', as indicated in the inset. $d_{sx}$ represents the skyrmion diameter. Inset: example of a snapshot of a Néel skyrmion stabilized by the i-DMI, where the arrows indicate the in-plane component of the magnetization while the colors are linked to the out-of-plane component (blue positive, red negative). (b) and (c) Skyrmion diameter as computed by means of micromagnetic simulations, corresponding to the distance from the geometrical center of the skyrmion to the region where $m_z$=0. (b) Skyrmion diameter as a function of $D$ for $k_U$=0.8 and 0.9MJ/m³ and (c) as a function of $k_U$ for $D$=3.0 and 3.5 mJ/m².

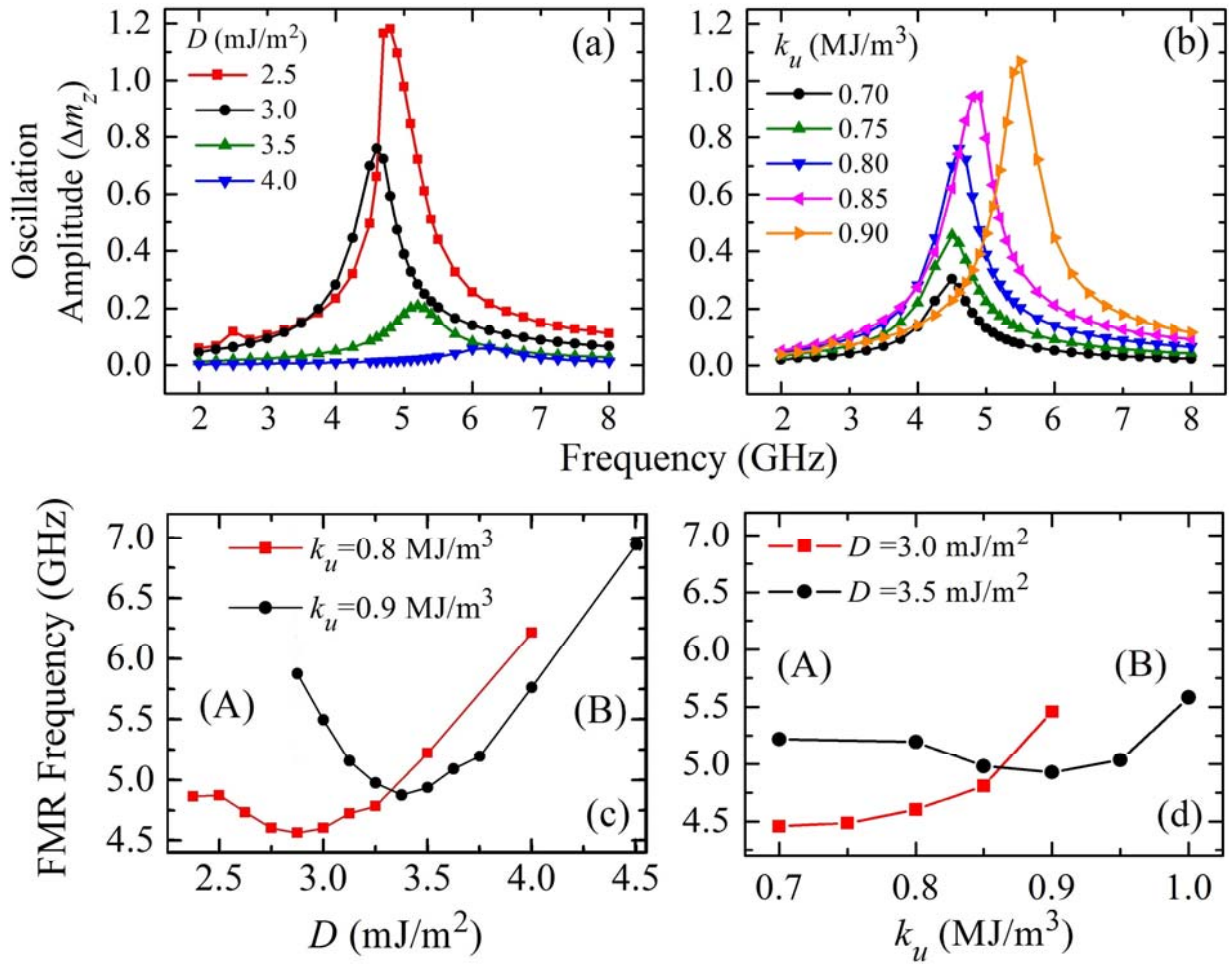

Figure 3 (a) STD response as a function of *D* for two different values of $k_U$. (b) STD response as a function of $k_U$ for two different values of *D*. (c) FMR frequency as a function of D for two different values of $k_U$. (d) FMR frequency as a function of $k_U$ for two different values of *D*.

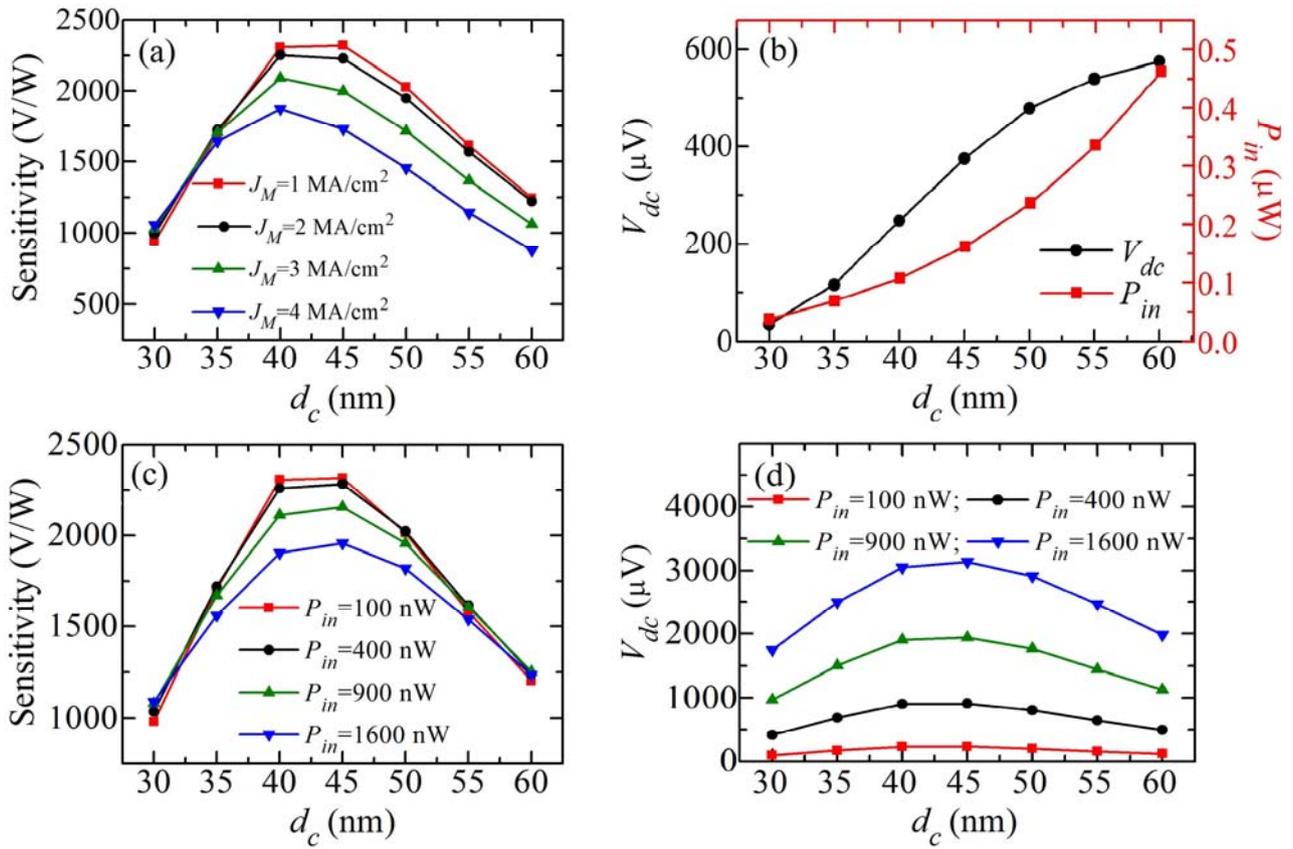

Figure 4(a) Sensitivity as a function of the contact diameter for different microwave currents as indicated in the main panels. (b) Detection voltage and microwave power as a function of the contact diameter for $J_M = 1\times10^6$ A/cm$^2$. (c) Sensitivity as a function of the contact diameter for different microwave powers as indicated in the main panels. In both (a) and (b) an optimal contact diameter corresponding to a maximum in the sensitivity can be observed. As can be observed the maximum does not coincide for computations at fixed current density and fixed microwave power. (d) Detection voltage as a function of the contact diameter for different microwave powers.

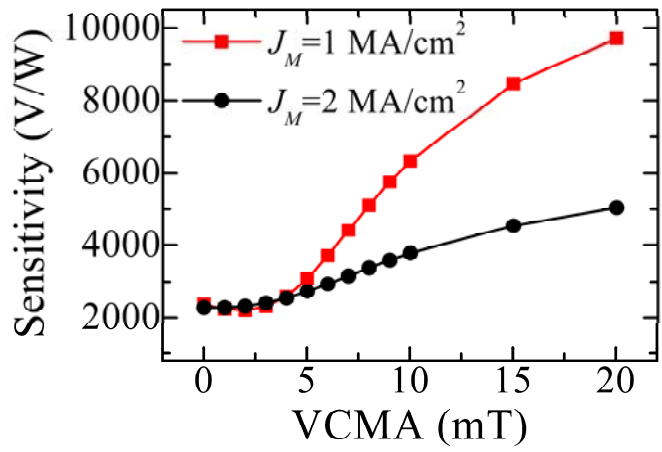

Figure 5. Sensitivity as a function of the VCMA, considering the optimal contact diameter of Fig. 4a, computed for $J_M$ =1 and 2x10$^6$A/cm$^2$.